\documentstyle[12pt,epsfig]{article}
 \newcommand{\be}{\begin{eqnarray}}
 \newcommand{\ee}{\end{eqnarray}}
 \newcommand{\beq}{\begin{equation}}
 \newcommand{\eeq}{\end{equation}}
 \newcommand{\ba}{\begin{array}{1}}
 \newcommand{\ea}{\end{array}}
 \newcommand{\bb}{\begin{thebibliography}}
 \newcommand{\eb}{\end{thebibliography}}

 \newcommand{\abstitle}[1]{{\small {\bf #1}}}
 \newcommand{\absauthor}[1]{{\small {\bf #1}}}
 \newcommand{\address}[1]{{\it #1}}
 
\begin{document}
 \begin{center}
 \abstitle{Search for intrinsic charm in vector boson production 
accompanied by heavy flavor jets}\\
 \vspace{0.6cm}
 \absauthor{P-H.Beauchemin$^1$, V.A.Bednyakov$^2$, G. I. Lykasov$^2$, 
 Yu.Yu. Stepanenko$^{2,3}$} 
\\ [0.6cm]
\address{$^1$ Tufts University, Medford, MA, USA\\
$^2$ Joint Institute for Nuclear Research -
  Dubna 141980, Moscow region, Russia\\
$^3$ Gomel State University, Gomel 246019, Republic of Belarus
}
 \end{center}
 \vspace{0.1cm}
 \vspace{0.2cm} 
\begin{center}
{\bf Abstract}
\end{center}
 \vspace{0.1cm} 
{
     Up to now, the existence of  intrinsic (or valence-like) heavy quark 
     component of the  proton distribution functions has not yet been confirmed or rejected. 
     The LHC with $pp$-collisions at $\sqrt{s}=$~7--13~TeV can supply us with extra unique 
     information concerning this hypothesis. 
     On the basis of our theoretical studies, it is demonstrated
     that investigations of the intrinsic heavy quark contributions
     look very promising in processes like $pp\rightarrow Z/W +c(b)+X$.
     A ratio of $Z+$ heavy jets over $W+$ heavy jets differential cross setion as a function of
     the leading jet transverse momentum is proposed to maximize the sensitivity to the intrinsic 
     charm component of the proton.
}

\section{Introduction}

     Parton distribution functions (PDFs) give the probability of finding in a proton a quark or a gluon 
     (parton) with a certain longitudinal momentum fraction at a given resolution scale.  The PDF 
     $f_a(x,\mu)$ is thus a function of the proton momentum fraction $x$ carried by the parton $a$ at the 
     QCD scale $\mu$. For small values of $\mu$, corresponding to long distance scales of less than 
     $1/\mu_0$, the PDFs cannot be calculated from the first principles of QCD (although some 
     progresses have been made using the lattice methods~\cite{LATTICE}). The unknown functions 
     $f_a(x,\mu_0)$ must be found empirically from a phenomenological model fitted to a large 
     variety of data at $\mu>\mu_0$ in a "QCD global analysis'' \cite{QCD_anal1,QCD_anal2}. The PDF 
     $f_a(x,\mu)$ at higher resolution scale $\mu>\mu_0$ can however be calculated from $f_a(x,\mu_0)$
     within the perturbative QCD using DGLAP $Q^2$-evolution equations~\cite{DGLAP}.  
     
     The limitation in the accuracy at which PDFs are determined constitutes an important source of 
     systematic uncertainty for Standard Model measurements and for multiple searches for New 
     Physics at hadron colliders. The LHC facility is a laboratory where PDFs can be studied and their 
     description improved. Inclusive $W^{\pm}$ and $Z$-boson production measurements performed 
     with the ATLAS detector have, for example, introduced a novel sensitivity to the strange quark density 
     at $x\sim0.01$~\cite{atlas-s}.

     Many $pp$ processes studied at LHC, including Higgs boson production, are sensitive to the 
     strange $f_s(x,\mu)$, charm $f_c(x,\mu)$, and/or bottom $f_b(x,\mu)$ quark distributions.  Global 
     analyses usually assume that the heavy flavor content of the proton at $\mu\sim m_{s,c,b}$ is 
     negligible. Theses heavy quark components arise only perturbatively through gluon-splitting as 
     described by the DGLAP $Q^2$-evolution equations~\cite{DGLAP}. Direct measurements of the open 
     charm, open bottom and open strangeness in Deep Inelastic Scattering (DIS) experiments confirmed 
     the perturbative origin of heavy quark flavors~\cite{H1:2005}. However, the description of these  
     experimental data is  not sensitive to the heavy quark distribution at large $x$ ($x>$0.1). 

     Analyzing hadroproduction of the so-called leading hadrons, Brodsky et al.~\cite{Brodsky:1980pb, 
     Brodsky:1981} have postulated, about thirty years ago, the co-existence of an {\it extrinsic} and an 
     {\it intrinsic} contributions to the quark-gluon structure of the proton. The {\it extrinsic} (or ordinary) 
     quarks and gluons are generated on a short-time scale associated with large-transverse-momentum 
     processes. Their distribution functions satisfy the standard QCD evolution equations. On the contrary, 
     the {\it intrinsic} quark and gluon components are assumed to exist over a timescale which is 
     independent of any probed momentum transfer. They can be associated with a bound-state 
     {(zero-momentum transfer regime)} hadron dynamics and one believes that they have a non-perturbative 
     origin. It was argued in~\cite{Brodsky:1981} that the existence of {\it intrinsic} heavy quark pairs 
     $c{\bar c}$ and $b{\bar b}$ within the proton state can be due to the virtue of gluon-exchange and 
     vacuum-polarization graphs. 
     
     A few models have been developed on this basis. The total probability of finding a quark from 
     the postulated intrinsic component of the PDF varies with these models. For example, in the MIT bag 
     model~\cite{Golowich:1981}, the probability of finding a five-quark component $|uudc{\bar c}\rangle$ 
     bounded within the nucleon bag, to which is associated an intrinsic charm component to the PDF, can 
     be of about 1--2\%. Another model consider a quasi-two-body state 
     ${\bar D}^0(u{\bar c})\, {\bar\Lambda}_c^+(udc)$ in the proton~\cite{Pumplin:2005yf}. In this scenario,
     the contribution of intrinsic charm ({\it IC}) to the proton PDF (the weight of the relevant Fock state in the 
     proton) could be as high as 3.5\%, with the upper limitation being due to constraints from DIS HERA 
     data~\cite{Pumplin:2005yf}--\cite{Nadolsky:2008zw}. In these models, the probability of finding an 
     intrinsic bottom state ({\it IB}) in the proton can also be estimated, but is suppressed by a factor of 
     $m^2_c/m^2_b\simeq 0.1$~\cite{Polyakov:1998rb} compared to intrinsic charm, where $m_c$ 
     and $m_b$ are respectively the masses of the charm quark ($\simeq$ 1.3 GeV) and of the bottom 
     quark (4.2 GeV). 
          
     It was recently shown that the possible existence of intrinsic strangeness in the proton results in a
     rather satisfactorily description of the HERMES data on $x(f_s(x,Q^2)+f_{\bar s}(x,Q^2))$ at 
     $x>$0.1 and $Q^2=$2.5 (GeV$/$c)$^2$ \cite{Peng_Chang:2012,LBDS:2013}. To further test the viability of
     the intrinsic quark hypothesis, intrinsic charm and bottom contributions must independently be 
     observed in other processes at other experiments. The typical high-$Q^2$ energy transfer yielded 
     in processes produced in recent high-energy hadron colliders might limit the observability of such 
     phenomena. The percent-level estimations for intrinsic charm contribution to PDF obtained from the 
     theoretical predictions quoted above have been calculated at low $Q^2$ and are decreasing as 
     $Q^2$ increases. Recent studies however showed that in high-energy LHC processes where a 
     charm-quark in the initial state leads to an heavy quark in the final state, intrinsic charm contribution 
     to the PDF could lead to an enhanced fraction of heavy meson (e.g. D-meson) in the final state 
     compared to when intrinsic quarks are ignored~\cite{LBPZ:2012}. This fraction is not independent of the phase space 
     probed. It was for example shown that selecting high-rapidity and large transverse momentum heavy 
     flavored jets enhances the $x>0.1$ PDF contribution to the cross section in the selected phase space, 
     and thus the intrinsic charm contribution to the observable number of events~\cite{LBPZ:2012}. 
     Experimental searches for a possible intrinsic charm signal at high-energy hadron collider like the LHC 
     are therefore possible. 
     
     Photons produced in association with heavy quarks $Q(\equiv c,b)$ in the final state of $pp$ collision 
     constitute a promising signature to look for intrinsic charm because, in contrary to dijet events where the 
     final state is dominated by gluon-jets or by a gluon splitting into a pair of heavy quarks, many diagrams 
     contain an heavy quark in both the initial and the final states~\cite{Brodsky:1980pb}-\cite{Nadolsky:2008zw}.
     Investigations of prompt photon and $c(b)$-jet production in $p{\bar p}$ collisions at $\sqrt{s}=1.96$~TeV 
     have been carried out at the TEVATRON~\cite{D0:2009}-\cite{CDF:2010}. An excess of $\gamma+c$ 
     events was observed over Standard Model expectations. In particular, it was observed that the ratio 
     of the experimental spectrum of a prompt photon accompanied by a $c$-jet to the relevant theoretical 
     expectation based on the conventional PDF ignoring {\it intrinsic} charm monotonically increases with 
     $p_T^\gamma$ up to a factor of about 3 when $p_T^\gamma$ reaches 110 GeV$/c$. While such trend 
     is expected from an intrinsic charm contribution to the PDFs, the magnitude of the effect observed in 
     the data is too large. Taking into account the CTEQ66c PDF, which includes the {\it IC} contribution 
     obtained within the BHPS model~\cite{Brodsky:1980pb,Brodsky:1981} leaves an excess of about 1.5
     times the new predictions~\cite{Stavreva:2009vi}. First studies of the $\gamma+b$-jets $p{\bar p}$-production 
     featured no enhancement in the $p_T^\gamma$-spectrum~\cite{D0:2009,CDF:2010}. A new version of
     such analysis published by the D\O\ collaboration in 2012 however presents the observation of such an 
     enhancement~\cite{D0:2012}, potentially hinting for an intrinsic bottom contribution to the PDFs. Because 
     of the ambiguity in these results, it becomes imperative to look for intrinsic quarks at the LHC.  

      Sensitivity studies of an {\it IC} signal in $p p\rightarrow \gamma+c(b)+X$ processes at LHC 
      energies was recently done in~\cite{BDLST:2014}. Results indicate that the possible existence of an 
      intrinsic heavy quark component in the proton can be inferred from the measurement of an 
      enhancement (by factor of 2 or 3) of the number of events with a photon of $p^\gamma_{T}>$ 150 GeV$/$c 
      at high rapidity $y_\gamma$ in comparison with the relevant number of events expected in the absence of  
      an {\it IC} contribution. The problem with $pp\rightarrow \gamma+c(b)+X$ processes 
      is that it is experimentally difficult to separate the prompt photon contribution from the non-prompt photon one,
      therefore adding ambiguities in the results of a measurement.

      A similar ${\it IC}$ signal can also be visible in the hard $pp$ processes of vector bosons production
      accompanied by heavy flavor ($b$ and $c$) jets in certain kinematic regions. These processes
      do not feature the problem mentioned above regarding $\gamma+c(b)+X$ processes. 
      In this paper, we study this heavy flavor production in $pp$ collisions at the LHC energies
      and discuss the potential observation of an intrinsic charm signal. In particular, we consider the production
      of the vector bosons $W^\pm, Z^0$ accompanied by $b(c)$-jets at large transverse momenta and propose
      a set of measurements to be performed at the LHC that maximize the sensitivity to the intrinsic 
      quark contribution to the PDFs.


\section{Intrinsic charm and beauty contribution to protons in W/Z plus heavy flavor jets at the LHC}
      
\begin{figure}[h!] 
\begin{center}
\begin{tabular}{cc}
\epsfig{file=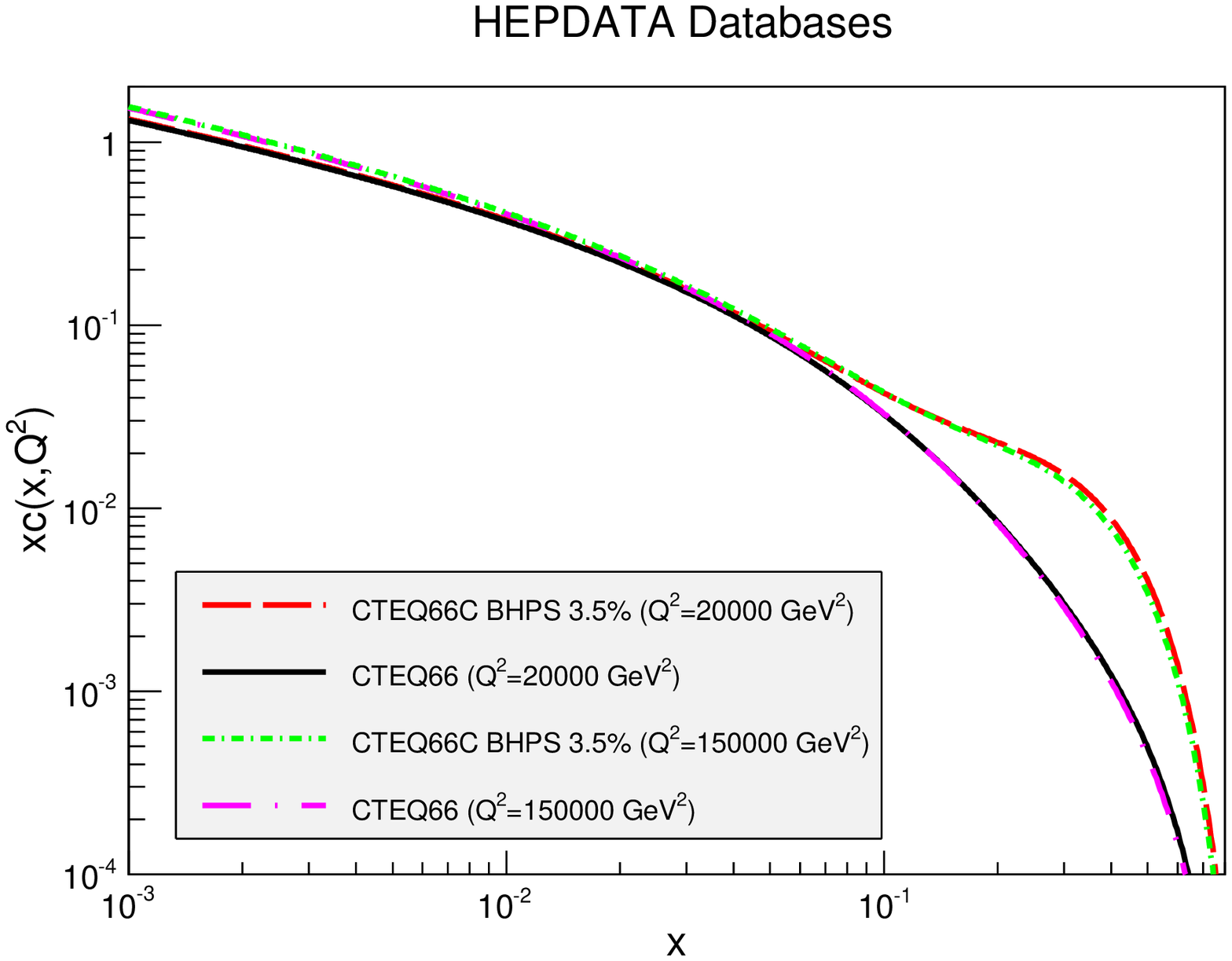,width=0.80\linewidth}
\end{tabular}
\end{center}
\caption{Distributions of the charm quark in the proton.
      	      The solid line is the standard perturbative sea charm density distribution $xc_{\rm rg}(x)$ at 
      	       $Q^2=20 000$ GeV $^2$, while the long dashed line is for $Q^2=150 000$ GeV $^2$.  
     	       The dashed curve corresponds to the charm quark distribution function for the sum of the intrinsic charm density 
      	       $xc_{\rm in}(x)$ and $xc_{\rm rg}(x)$ at $Q^2=20 000$ GeV $^2$, while the short dashed line is for 
     	       $Q^2=150 000$ GeV $^2$ \cite{Nadolsky:2008zw}.} 
\label{Fig_1IC}
\end{figure}

      At the LHC, with a center of mass energy of $\sqrt{s}=$7-13 TeV, the typical momentum transfer  
      squared ($Q^2$) in the hard $pp$ processes of photon or vector boson production accompanied   
      by heavy flavor jets with large transverse momenta is above a few tenth of thousands of  
      (GeV/c)$^2$. At these scales, the contribution of the intrinsic charm component to the PDF 
      features an enhancement at $x\ge$ 0.1, where the corresponding PDFs turn out to be larger  
      (by more than an order of magnitude at $x\sim$ 0.3-0.4) than the sea ({\it extrinsic}) charm density  
      distribution in the proton~\cite{BDLST:2014,Nadolsky:2008zw}. This can be seen in Fig.~\ref{Fig_1IC}  
      where we compare the charm distribution in proton for two values of $Q^2$ ($Q^2=$ 20 000 
      (GeV/c)$^2$ and $Q^2=$ 150 000 (GeV/c)$^2$) for a PDF, CTEQ66c, which includes an {\it IC} 
      component, and another one, CTEQ66, including only the extrinsic quark contribution. From  
      the figure, one can also see that the high-$Q^2$ dependence of $xc(x,Q^2)$ does not affect much the 
      PDF values at $x>0.1$.
      
      The sensitivity studies performed with $\gamma+c(b)+X$ processes in the context of the LHC 
      demonstrated that with the appropriate phase space selections on the final state photon and heavy 
      flavor jet, one can select a large fraction of events with $x_c>$0.1, and thus substantially intensifies 
      the intrinsic charm PDF contribution to charm hadroproduction, enough for being able to see a 
      signal when compared to extrinsic contribution only~\cite{BDLST:2014}.
 
{     In particular, it was shown that in at least one of the colliding protons the momentum fraction
      of $c$-quark $x_c$ should be not less than the Feynman variable of photon $x_F^\gamma$, i.e.,
\be
x_c\geq x_F^\gamma=\frac{2p_T^\gamma}{\sqrt{s}sinh(\eta_\gamma)}~,
\label{def:xcxF}    
\ee      
      where $p_T^\gamma$ is the transverse momentum of photon and $\eta_\gamma$ is its pseudo-rapidity.
      Therefore, at $x_F^\gamma\geq 0.1$, or equivalently 
      at the charm momentum fraction $x_c> 0.1$ 
      the {\it intrinsic} charm distribution substantially intensifies the charm 
      PDF contribution to the process $pp\rightarrow\gamma+c+X$ 
(see Fig.~\ref{Fig_1IC}). 
      As a result, 
      the $p_T^\gamma$-spectrum will feature a significant enhancement with
      respect to the expected spectrum with non-intrinsic charm contribution 
      in that region of the $p_T^\gamma$, $\eta_\gamma$ phase space where $x_F^\gamma\geq 0.1$
      in accordance to (\ref{def:xcxF}).

      Such region can be selected, for example, by requiring a prompt photon and a final state $c$-jet with rapidity 
      close to the pseudo-rapidity $\eta_\gamma$ at large $p_T^\gamma$   
      of respectively  
      1.5$<\mid y_\gamma\mid<$ 2.4 and $\mid y_c\mid<2.4$, and by imposing large transverse momenta 
      cuts ($>150$ GeV) to the photon and to the heavy flavor jet. The observation of an excess of events 
      selected in this phase space region compared to the standard non-{\it IC} component would thus
      provide a compiling evidence for the existence of intrinsic charm, and could be use to estimate
      the increase in the PDF due to intrinsic charm as a function of $x$. 

      This strategy can equally be applied to test and measure the intrinsic heavy quark contribution to
      the production of vector bosons  $W^\pm,Z^0$ accompanied by heavy flavor jets ($Q_f$-jets,
      with $Q_f=s,c,b$). In these events, the intrinsic quark component would receive its main contribution
      from $Q_f({\bar Q}_f)+g\rightarrow W^\pm/Z^0+Q_f^\prime({\bar Q}_f^\prime)/Q_f({\bar Q}_f)$ processes for
      which the LO QCD diagrams are presented in Figs.~\ref{Fig_2Fd} and~\ref{Fig_3Fd}, in the case of the
      Z and W production, respectively. Here $Q_f^\prime=c,s,c$ if $Q_f=s,c,b$, respectively.
  
      At NLO in QCD, $W/Z+Q_f$ diagrams, often more complicated than the ones presented in
      Figs.~(\ref{Fig_2Fd},\ref{Fig_3Fd}), must also be considered. As can be seen in Fig.~\ref{Fig_6Wb}, 
      the heavy flavor jets in the final state of these diagrams come from a gluon splitting somewhere along 
      the event chain, and does thus not feature any intrinsic quark contribution. If the cross sections of these
      diagrams is large enough, the conclusions about the sensitivity of a measurement to intrinsic charm
      will be affected. It is thus important to consider QCD NLO calculations in the current study. 
      
      To this end, we calculated the $p_T$-spectra of heavy flavor jets ($b$ and $c$) in association with a vector 
      boson produced at NLO in $pp$ collisions at $\sqrt{s}=$8, 13 TeV using the parton level Monte Carlo (MC) 
      generator MCFM version 6.7 \cite{MCFM}. For the various processes considered, the vector boson is required 
      to decay leptonically, in order to allow experimental studied to trigger on these events, and the pseudo-rapidity 
      of the heavy quark jet is required to satisfy $\eta >$ 1.52, to probe high-$x$ PDFs.

      By selecting $Z+c$-jet events, where the $c$-jet is required to be rather forward (1.5$<\mid y_c\mid<$ 2.0), 
      we can see on the left panel of Fig.~\ref{Fig_4Zc} that the $c$-jet transverse momentum spectrum of events 
      with a 3.5\% intrinsic charm contribution to the PDF (CTEQ66c) features an excess, increasing with the 
      $c$-jet $p_T$, compared to the corresponding differential cross section when only extrinsic components 
      of the PDF are considered (CTEQ66). These differential cross section distributions have been obtained 
      at NLO from the MCFM process 262. From the right panel of the same figure, showing the ratio of the two 
      spectra obtained with and without {\it IC} contribution, we can see that the excess in the $c$-jet $p_T$ 
      spectrum due to {\it IC} is of $\sim5$\% for $p_T$ of 50 GeV, and rises to about 220$\%$ for $p_T\sim300$ 
      GeV. This effect can thus be observed at the LHC if the $c$-jet $p_T$ differential cross section in $Z+c$ 
      events can be measured with sufficient precision. 
      
      In the case of the W production in association with heavy flavor jets, the intrinsic charm contribution would
      be observed in a $W+b$-jet final state due to the change of flavor in the charged current. In MCFM, the NLO 
      $W+b$ Feynman diagrams for which the LO part is depicted in Fig.~\ref{Fig_3Fd}, correspond to the processes 
      12 and 17~\cite{MCFM}. They provide the contribution to $W+Q'$  which is sensitive to {\it IC}. The $p_T$ 
      spectrum of the b-jet is presented, for the sum of these processes, in Fig.~\ref{Fig_5Wb} (left), where one 
      calculation (squares) has been obtained at NLO in QCD with the CTEQ66c PDF that includes an {\it IC} 
      contribution (about 3.5$\%$), and the other calculation (triangles) uses the CTEQ66 PDF, which does not 
      include {\it IC}. On the right panel of Fig.~\ref{Fig_5Wb}, the ratio of these two spectra with and without an 
      {\it IC} contribution to the PDF used in the $W+b$ production calculations is presented. From this figure, 
      one can see that the inclusion of the {\it IC} contribution to the PDF leads to an increase in the $b$-jet 
      spectrum by a factor of about 2-2.5 at $p_T^{jet} >$ 250 GeV$/$c. This is comparable to what was 
      observed in the $Z+c$ case of Fig.~\ref{Fig_4Zc}.

\begin{figure}[h!]
\begin{center}
\epsfig{file=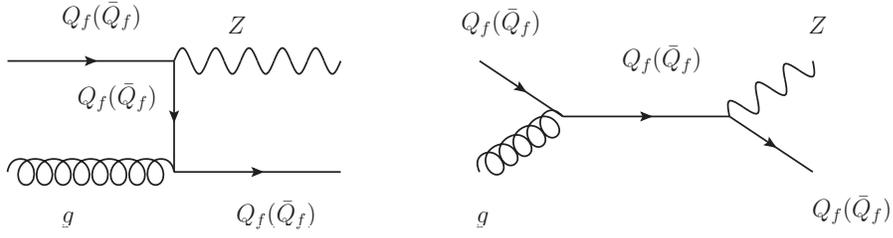,width=0.80\linewidth}
\end{center}
\caption{LO Feynman diagrams for the process $Q_f({\bar Q}_f) g\rightarrow Z Q_f({\bar Q}_f)$}
\label{Fig_2Fd}
\end{figure}

\begin{figure}[h!]
\begin{center}
\begin{tabular}{cc}
\hspace{0.5cm}%
\epsfig{file=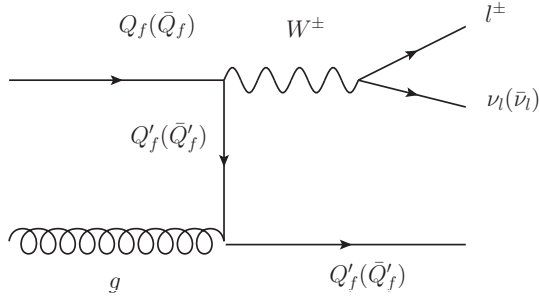,width=0.50\linewidth}
\end{tabular}
\end{center}
\caption{Example of an LO Feynman diagram for the process $Q_f({\bar Q}_f) g\rightarrow W^\pm Q_f^\prime({\bar Q}_f^\prime)$,
                where $Q_f=c,b$ and $Q_f^\prime=b,c$ respectively.}
\label{Fig_3Fd}
\end{figure}

\begin{figure}[h!]
\begin{center}
\begin{tabular}{cc}
\hspace{0.5cm}%
\mbox{\epsfig{file=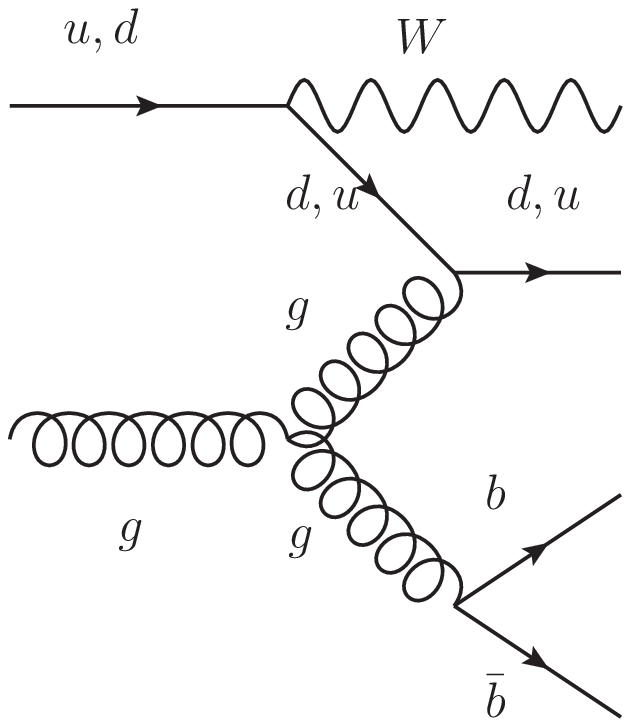,width=0.35\linewidth}
\epsfig{file=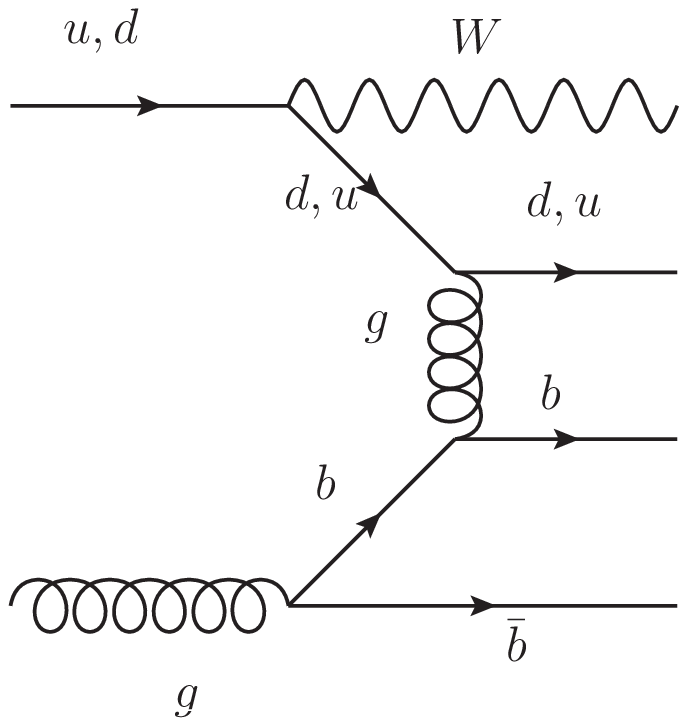,width=0.40\linewidth}}
\end{tabular}
\end{center}
\caption{Some NLO Feynman diagrams for the process $Q_f({\bar Q}_f) g\rightarrow W^\pm Q_f^\prime({\bar Q}_f^\prime)$,
	      where $Q_f=c,b$ and $Q_f^\prime=b,c$ respectively. Left: gluon-splitting; Right: $t$-channel type of 
	      W-scattering with one gluon exchange in the intermediate state.}
\label{Fig_6Wb}
\end{figure}

\begin{figure}[h!]
\begin{center}
\begin{tabular}{cc}
\hspace{0.5cm}
\mbox{\epsfig{file=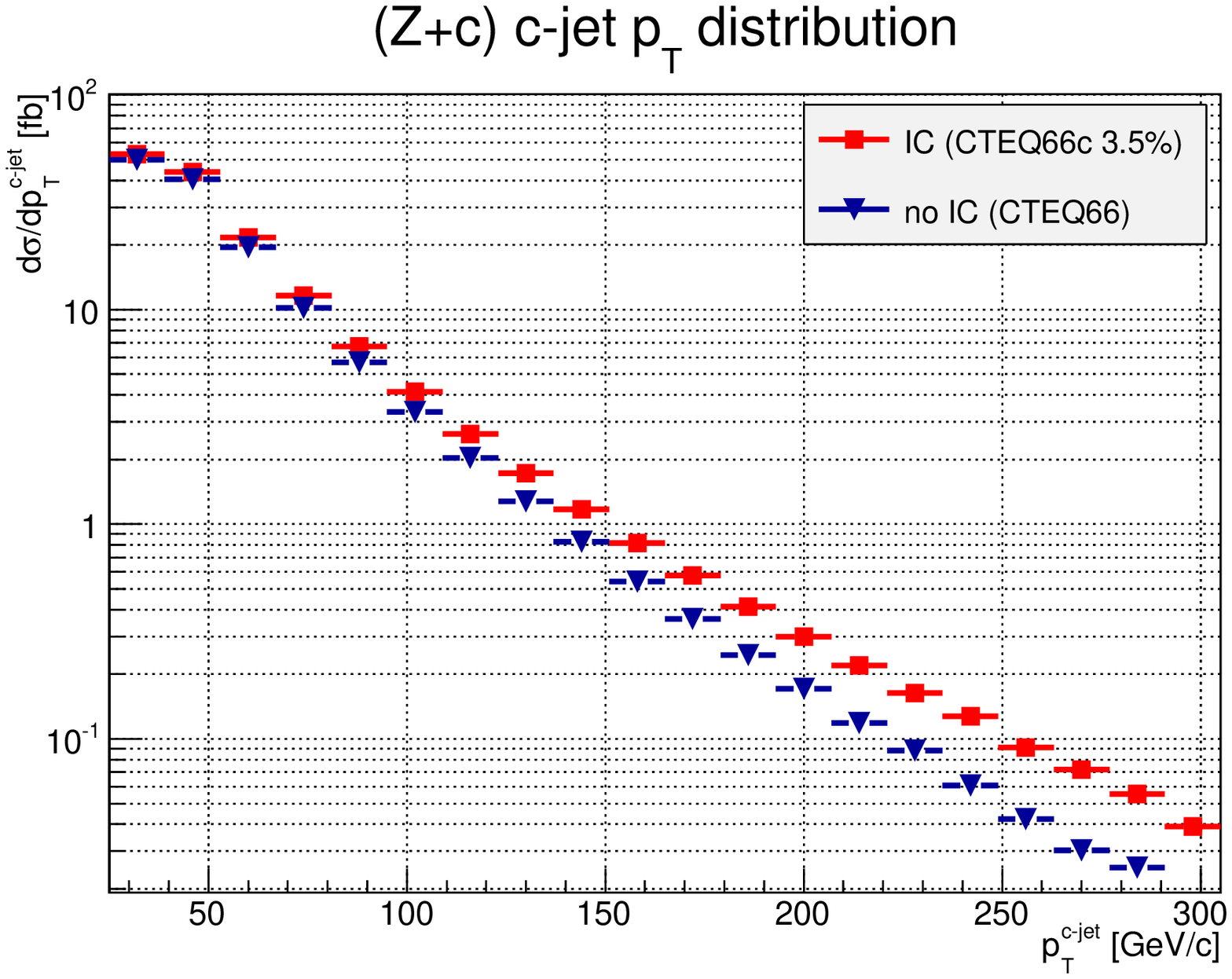,width=0.45\linewidth}
\epsfig{file=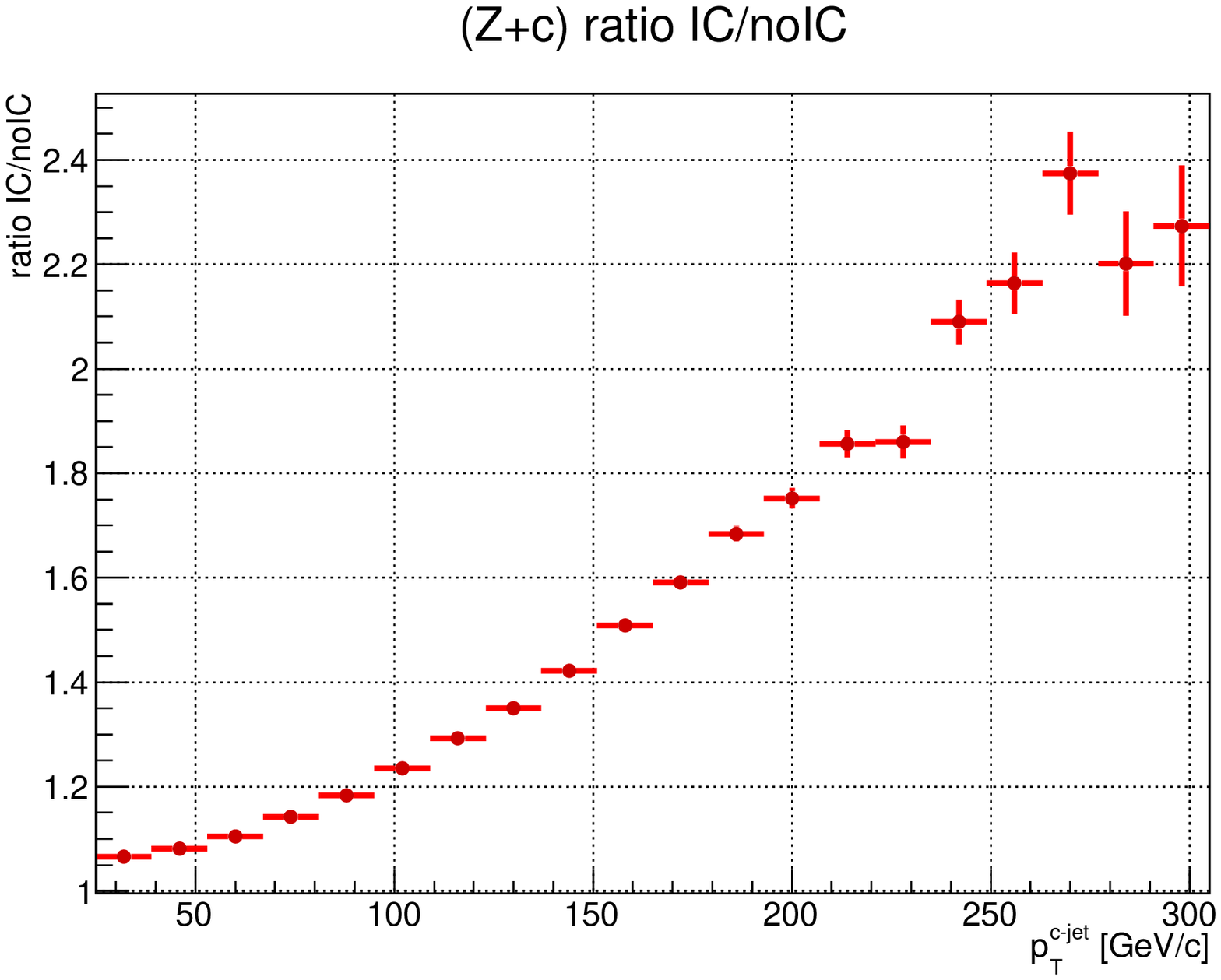,width=0.45\linewidth}}
\end{tabular}
\end{center}
\caption{Left: Comparison of the $p_T$-spectra for the NLO $pp\rightarrow Z+c$
               process 262 \cite{MCFM} obtained with
 	      PDF including an intrinsic charm component (CTEQ66c) and PDF having only an extrinsic component 
 	      (CTEQ66). Right: Ratio of these two spectra.}
\label{Fig_4Zc}
\end{figure}

\begin{figure}[h!]
\begin{center}
\begin{tabular}{cc}
\hspace{0.5cm}
\mbox{\epsfig{file=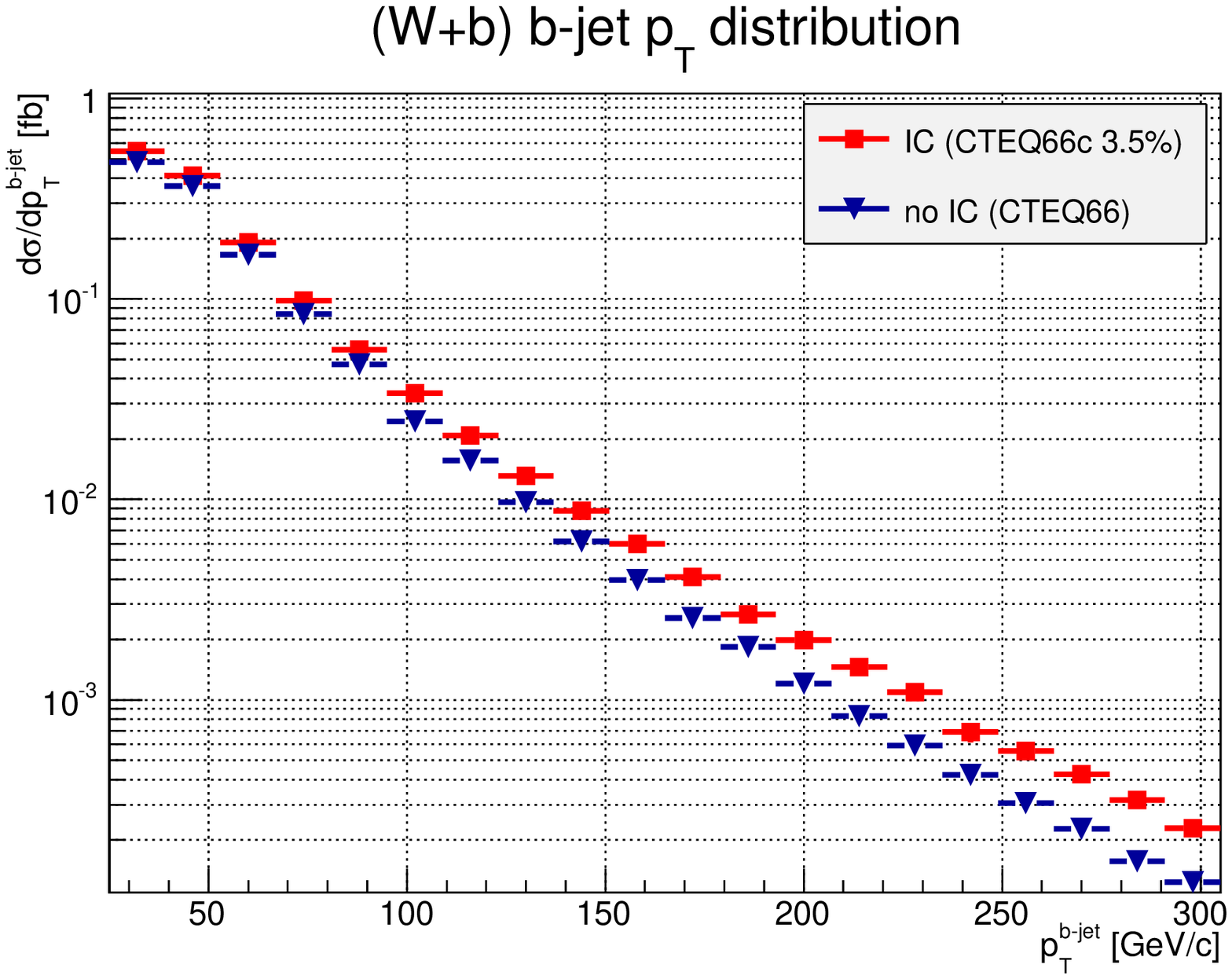,width=0.45\linewidth}
\epsfig{file=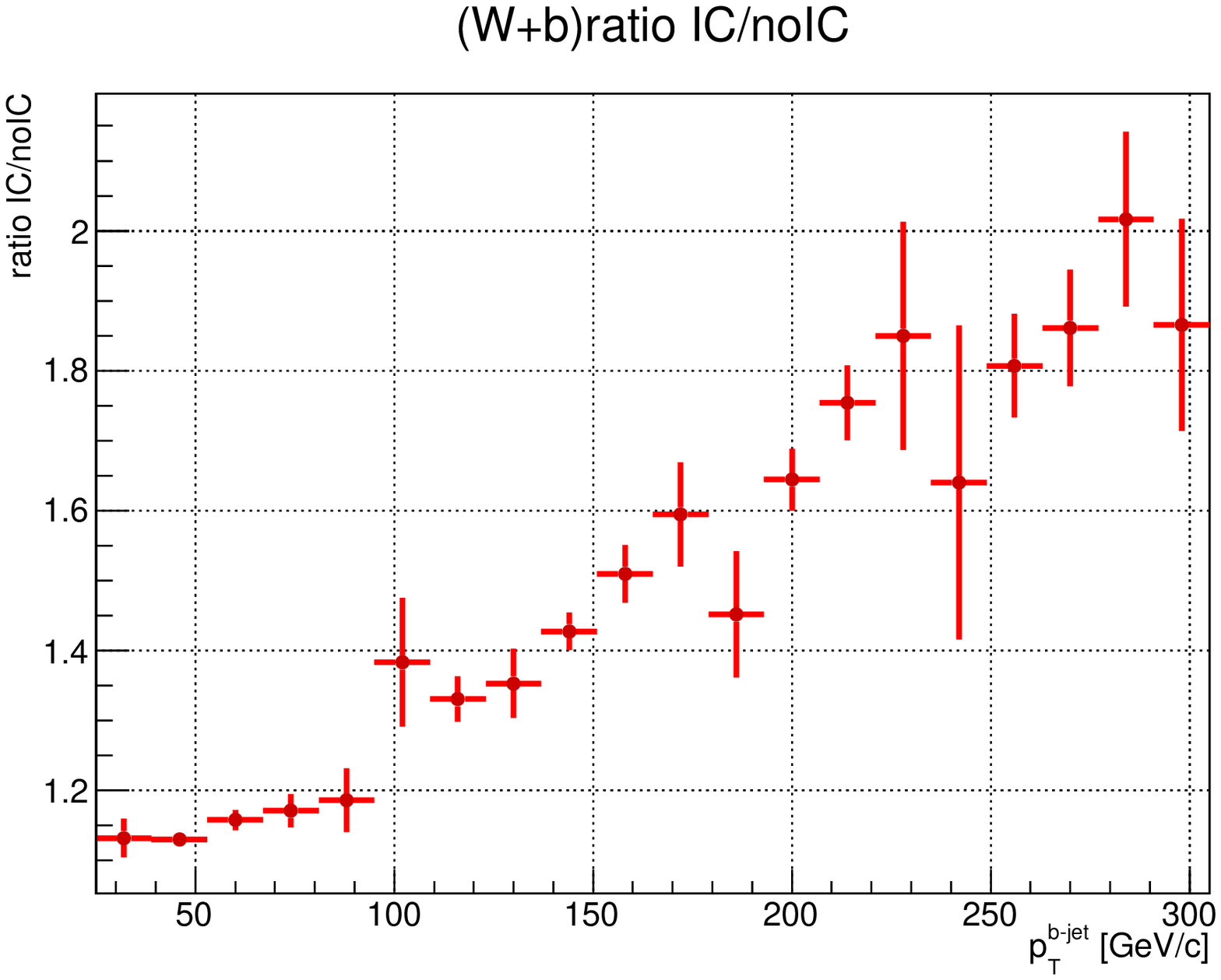,width=0.45\linewidth}}
\end{tabular}
\end{center}
\caption{Left: Comparison of the $p_T$-spectra for the NLO $pp\rightarrow W+b$,
               processes 12+17 \cite{MCFM} obtained with
               PDF including an intrinsic charm component (CTEQ66c) and PDF having only an extrinsic component 
               (CTEQ66). Right: Ratio of these two spectra.}
\label{Fig_5Wb}
\end{figure}

   Similarly, the $W+c$ final state would be sensitive to intrinsic strange while the $Z+b$ final state would be sensitive 
   to intrinsic bottom.  These processes are however suboptimal for finding intrinsic quarks at the LHC. As mentioned
   above, the contribution of the ${\it IB}$ to the PDF is suppressed by a factor of $(\frac{m_c}{m_b})^2$ and
   is thus subdominant compared to intrinsic charm. The contribution of the intrinsic strangeness (${\it IS}$) 
   can be of the same order of magnitude as the ${\it IC}$ according to~\cite{Peng_Chang:2012,LBDS:2013}. The $Q^2$
   evolution for this component has however not been calculated up to now, and thus contains many unknown. This
   is why this paper concentrate on the intrinsic charm component of the proton. 
                 
   The above results of Figs.~\ref{Fig_4Zc} and~\ref{Fig_5Wb} seem a priori very encouraging regarding the capacity 
   of the LHC to provide an observation of an intrinsic charm contribution to the PDFs in $W/Z+Q_f$ events, but the 
   real situation is unfortunately more complex than this. The $W$ boson plus one or more $b$-quark jets production,
   calculated at NLO in the 4-flavor scheme (4FNS), for which two of the diagrams are represented in Fig.~\ref{Fig_6Wb}
   must also be included. These corresponds to the MCFM processes 401/406 and 402/407~\cite{MCFM}. Their total cross 
   section is about 50 times larger that the $W+b$ processes sensitive to {\it IC}. As a result, the total $W+b$ production
   is not sensitive to an intrinsic charm component of the PDF, as can be seen in Fig.~\ref{Fig_5Wbj}, where the sum of
   all processes contributing to $W+b$ have been taken. Fortunately the $Z+c$ processes do not suffer from a similar 
   large dilution of the intrinsic quark component because the $Q_f+g\rightarrow Z+Q_f$ processes, which are sensitive 
   to {\it IC}, are not Cabibbo-suppressed.
           
   Another difficulty consists in the experimental identification of heavy flavor jets in order to select, for example, 
   $Z+c$-jet events in a very large $Z$+jets sample. Algorithms typically disentangling heavy flavor jets from 
   light-quark jets exploit the longer lifetime of heavy-quark hadrons that decay away from the primary 
   vertex of the main process, but close enough to allow for a reconstruction of the tracks of the decay 
   products of the heavy-flavor hadron in the inner part of the detector. Such algorithms are typically not 
   capable of explicitly distinguishing $c$-jets from $b$-jets; only the efficiency for identifying the heavy flavor
   nature of the jet would differ between $c$-jets and $b$-jets. For example, one of the ATLAS heavy-flavor 
   tagging algorithm (MV1) yields an efficiency of 85\% for $b$-jet identification and 50\% for $c$-jet (for a working 
   point where the light flavor rejection is 10)~\cite{ATLAS:13-109}. As a result of such heavy flavor jet 
   tagging algorithm, the selected $Z+Q$ final state will be a mixture of $Z+c$ and $Z+b$.  
   
   A priori, one would expect that $Z+b$ events are sensitive to intrinsic bottom and therefore act as a 
   background to intrinsic charm studies, when the two processes cannot be experimentally distinguished.
   The situation is however more complicated than this. Because of sum rules, an intrinsic charm component
   would affect the total b-quark contribution to the proton, and the $Z+b$-jet final state therefore becomes 
   sensitive to intrinsic charm as well. As can be seen in Fig.~\ref{Fig_8Zb}, this contribution is in the opposite 
   direction of the intrinsic charm effect on $Z+c$ processes presented in Fig.~\ref{Fig_4Zc}. In addition, the 
   heavy flavor tagging efficiency is lower for $c$-jets than it is for $b$-jets, therefore increasing the weight of 
   the negative $Z+b$ contribution to the total Z plus heavy flavor tagged jets signal. The question is thus: are 
   $Z+Q$-jet events still sensitive to intrinsic charm?

\begin{figure}[h!]
\begin{center}
\begin{tabular}{cc}
\hspace{0.5cm}
\mbox{\epsfig{file=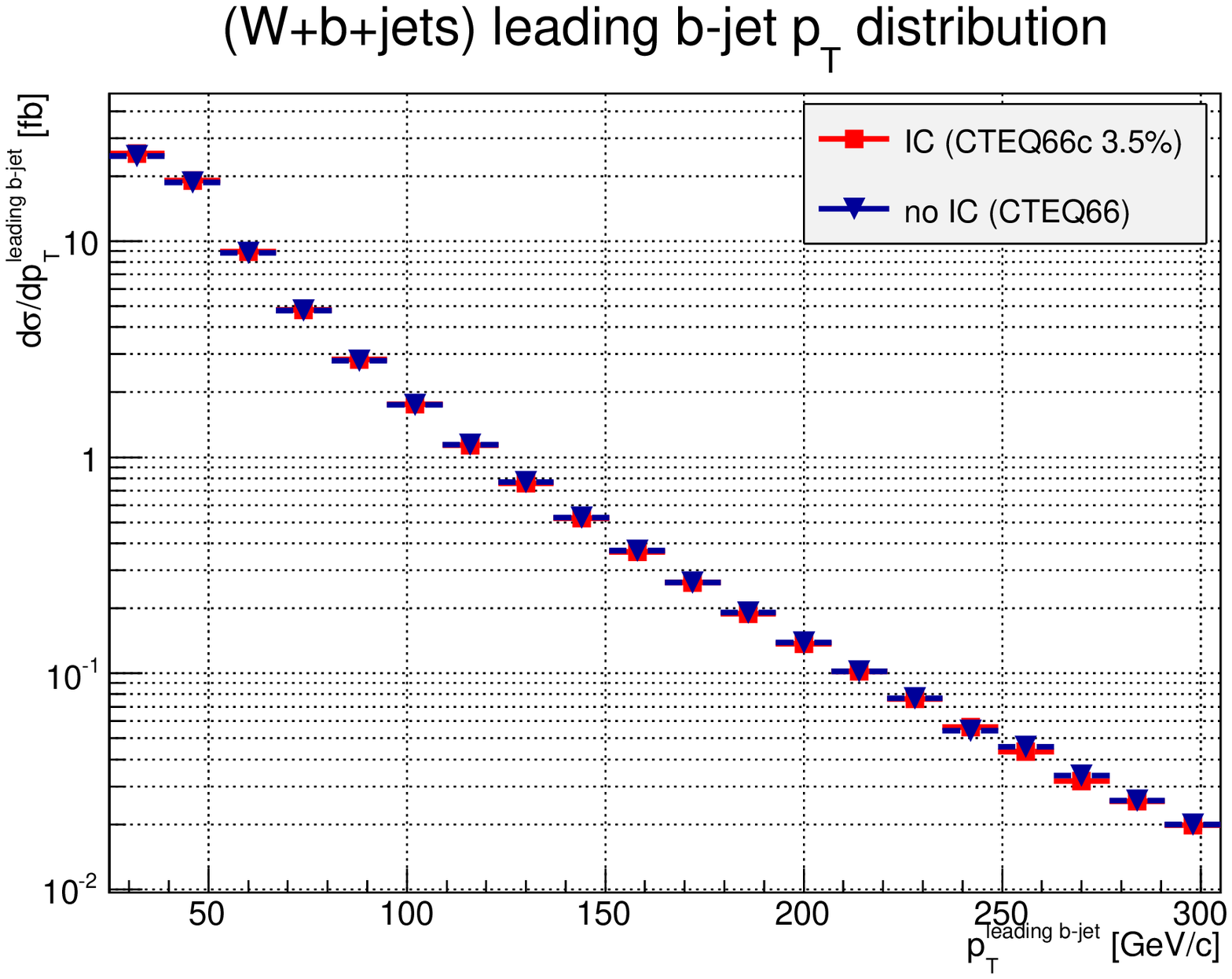,width=0.45\linewidth}
\epsfig{file=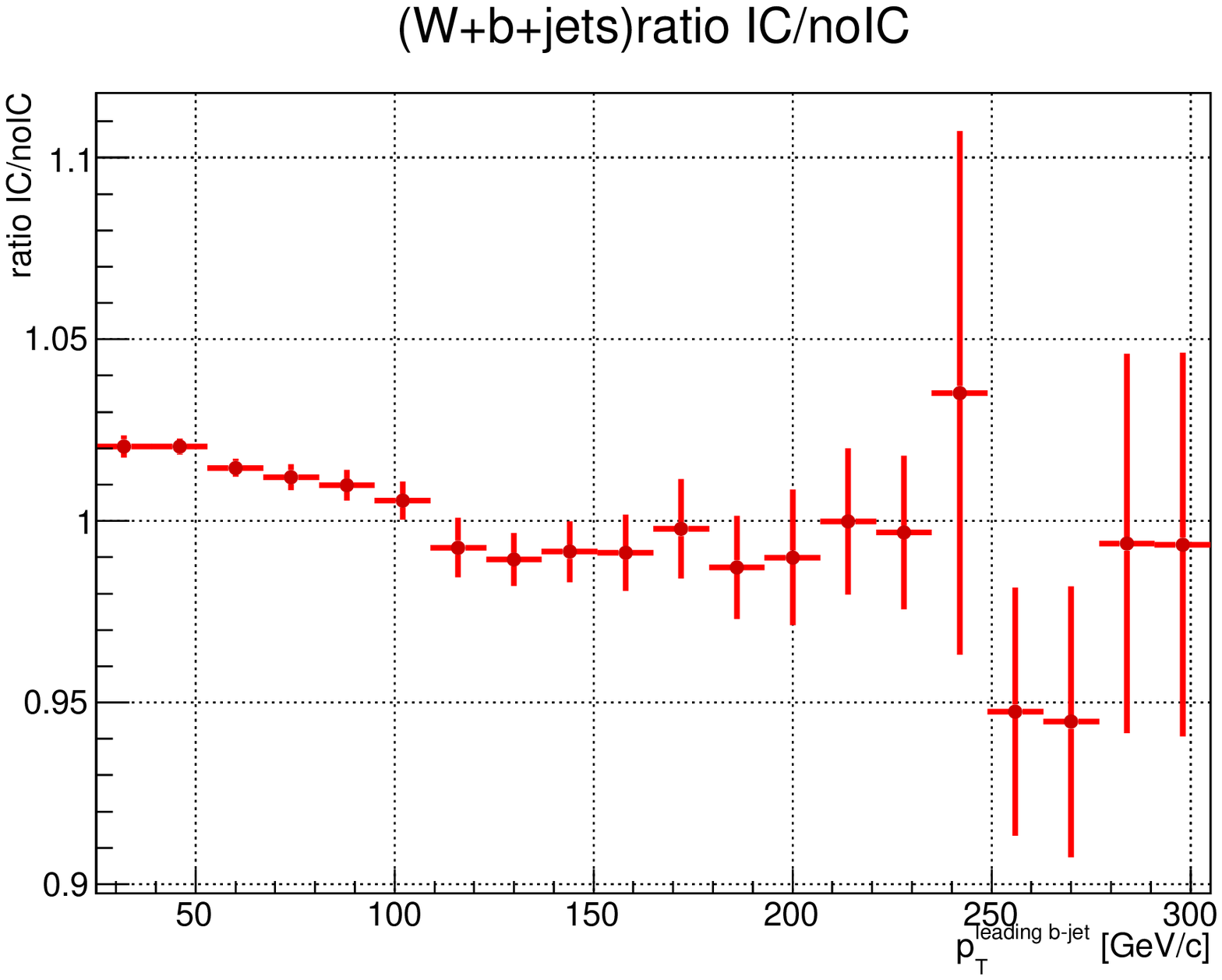,width=0.45\linewidth}}
\end{tabular}
\end{center}
\caption{Left: Comparison of the $p_T$-spectra for the NLO $pp\rightarrow W+b +jet$ processes 
               401/406 and 402/407 \cite{MCFM} obtained with
               PDF including an intrinsic charm component (CTEQ66c) and PDF having only an extrinsic component 
               (CTEQ66). Right: Ratio of these two spectra.}
\label{Fig_5Wbj}
\end{figure}

\begin{figure}[h!]
\begin{center}
\begin{tabular}{cc}
\hspace{0.5cm}
\mbox{\epsfig{file=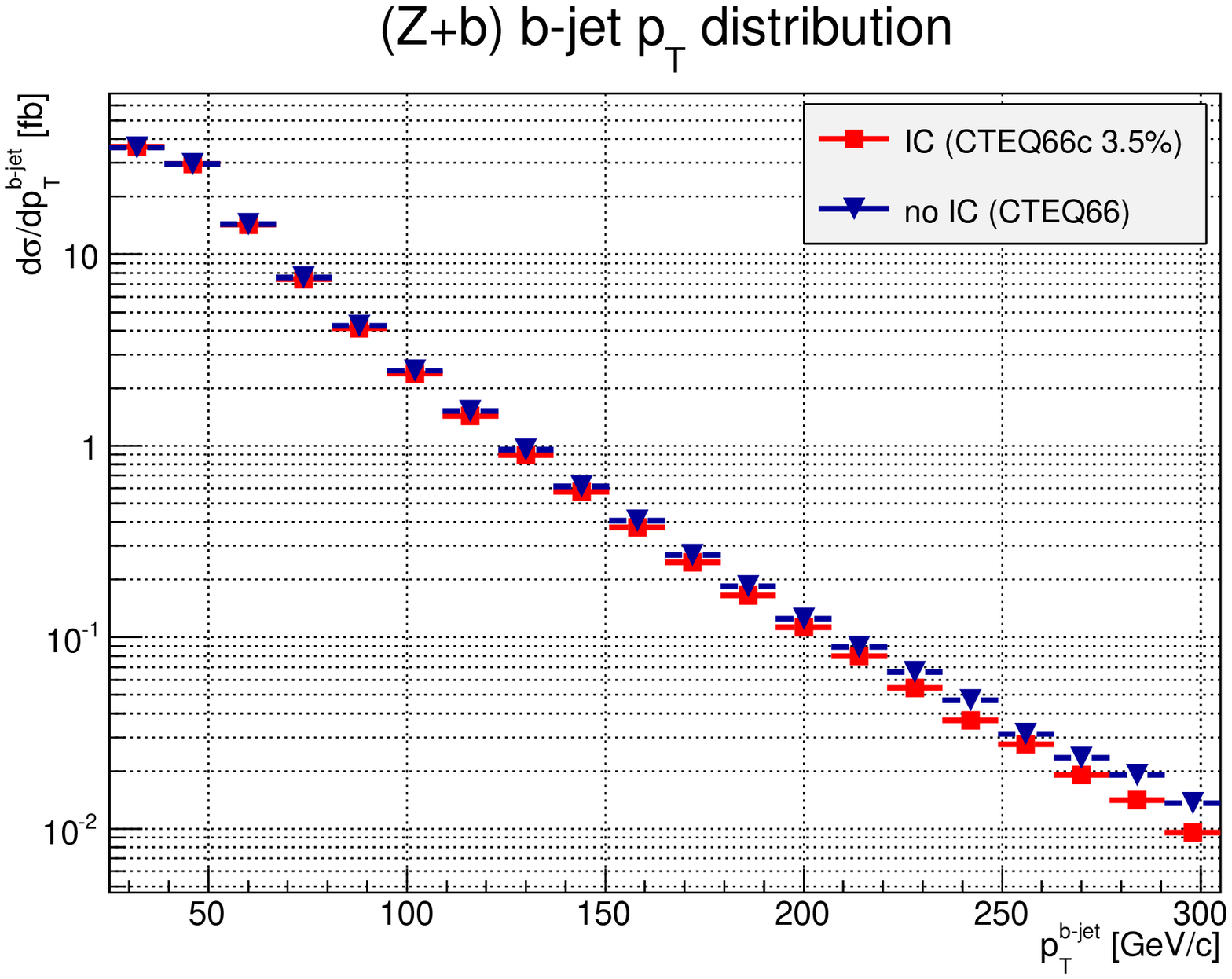,width=0.50\linewidth}
\epsfig{file=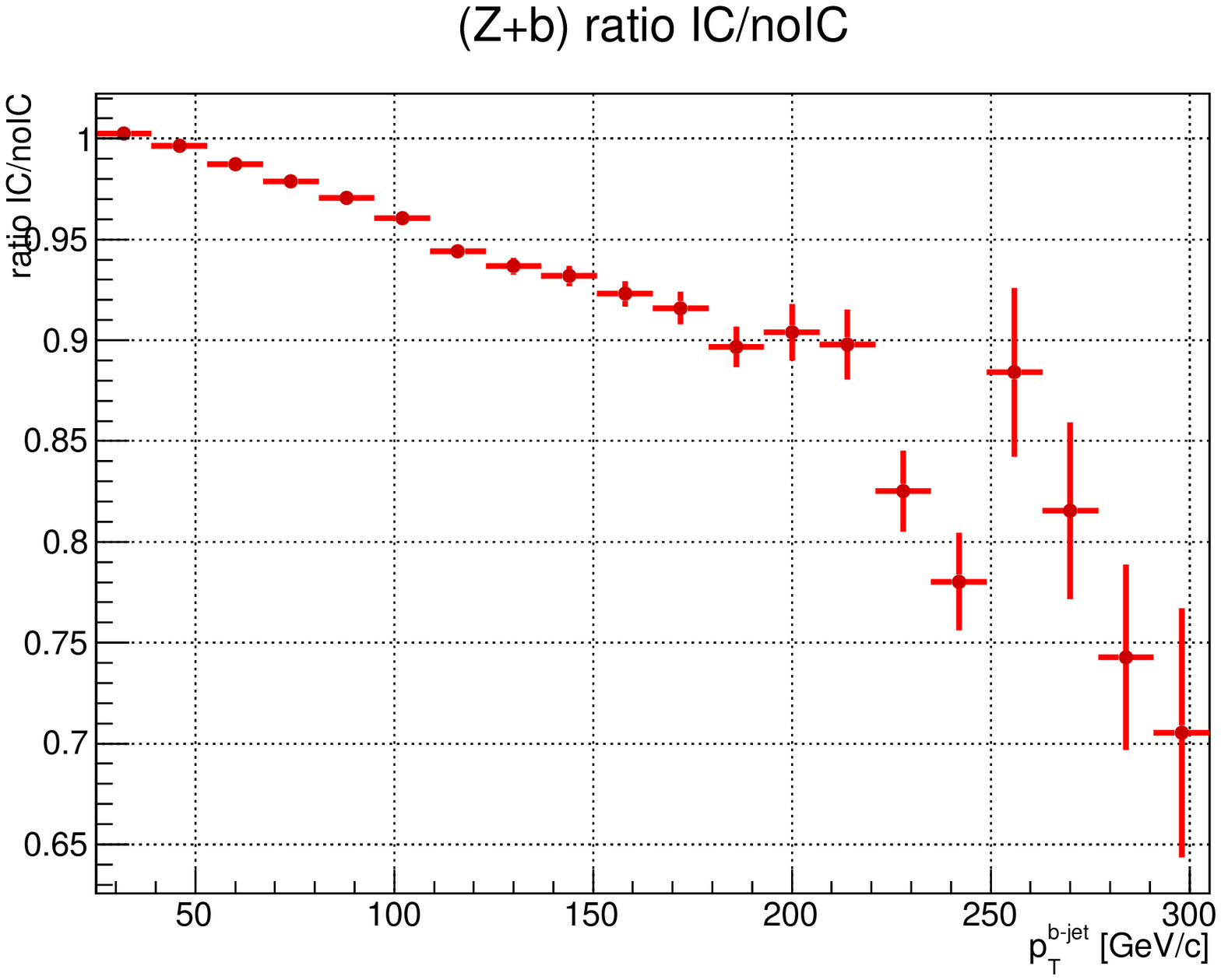,width=0.50\linewidth}}
\end{tabular}
\end{center}
\caption{Left: Comparison of the $p_T$-spectra for the total NLO $pp\rightarrow Z+b$ process 261 obtained
               with PDF including an intrinsic charm component (CTEQ66c) and PDF having only an extrinsic component 
	       (CTEQ66). Right: Ratio of these two spectra. }
\label{Fig_8Zb}
\end{figure}

   To answer this question, we once again used MCFM to calculate the $p_T$ spectra of the heavy 
   flavor jets at NLO for $Z+c$ (processes 262, 267) and for $Z+b$ (processes 261, 266)~\cite{MCFM}.
   This includes the contribution from both heavy flavor scattering and pair production from gluon-splitting. 
   In all processes, the Z-boson is required to decay leptonically, and a pseudo-rapidity
   cut of  $\eta >$ 1.52 is applied on the heavy flavor jets. We also applied the $b$-tagging 
   and $c$-jet tagging efficiency on the corresponding jet, as a function of the $p_T$ of the jet, 
   as reported in~\cite{ATLAS:13-109}. The resulting spectrum from all the processes has then been 
   summed. In Fig.~\ref{Fig_9Zall}, we can see that despite the negative contribution of $Z+b$ 
   processes and the effect of heavy flavor tagging efficiency, the contribution of intrinsic charm 
   has a significant impact on the shape and normalization of the heavy flavor jet spectrum, 
   suggesting that it can be tested at the LHC.

\begin{figure}[h!]
\begin{center}
\begin{tabular}{cc}
\hspace{0.5cm}
\mbox{\epsfig{file=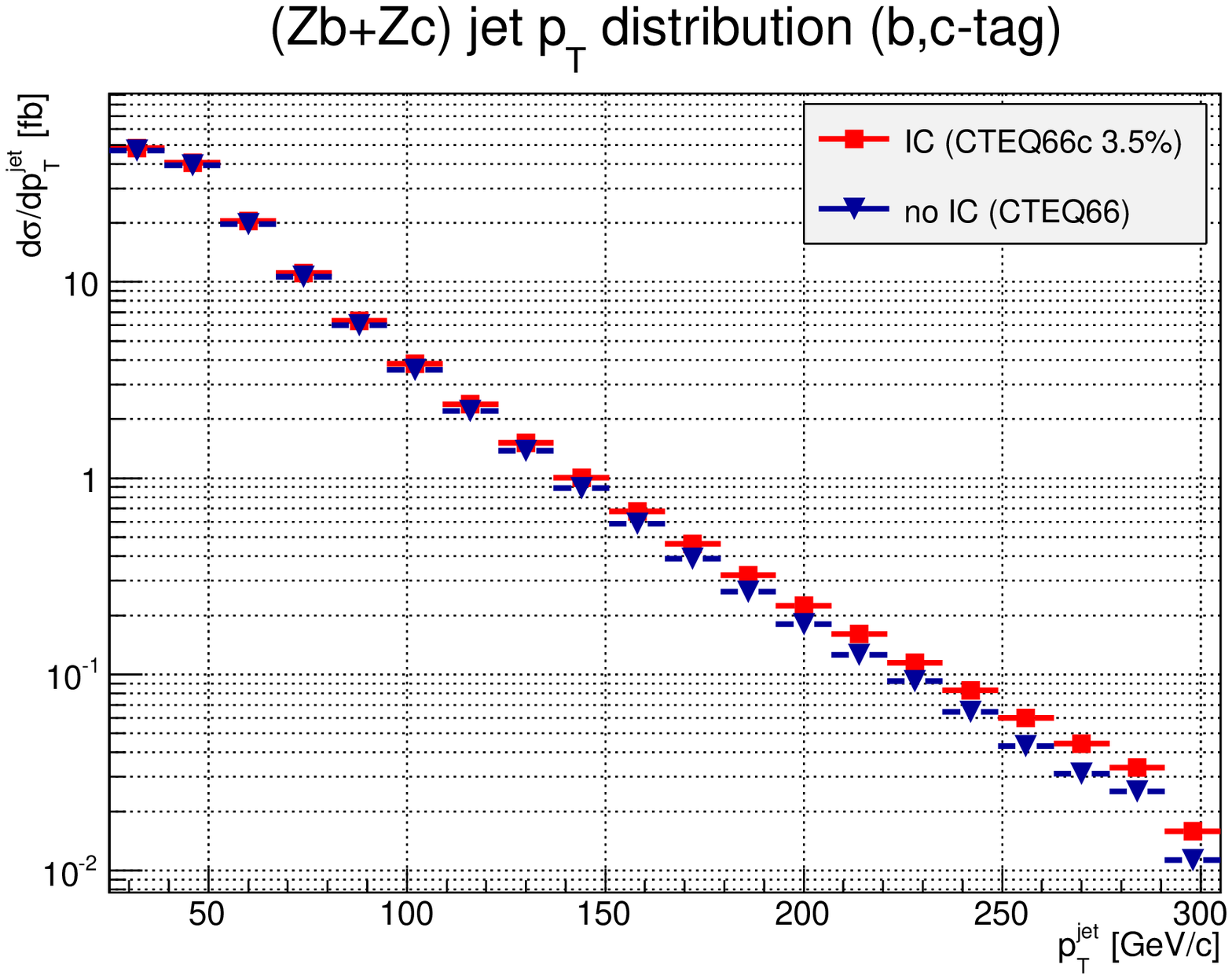,width=0.45\linewidth}
\epsfig{file=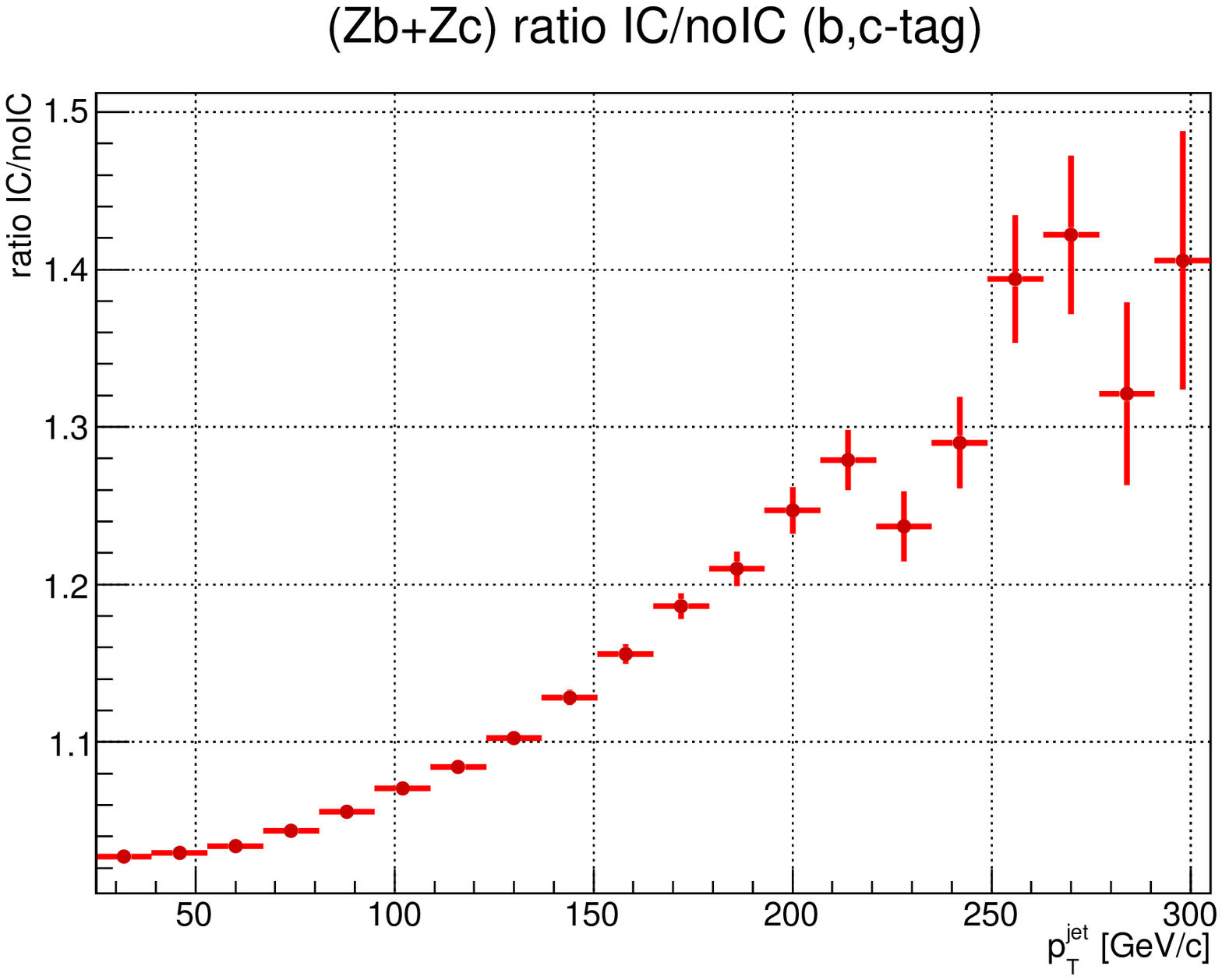,width=0.45\linewidth}}
\end{tabular}
\end{center}
\caption{Left: Comparison of the $p_T$-spectra for the total NLO $pp\rightarrow Z+b(\bar{b})$ 
               process plus $pp\rightarrow Z+c(\bar{c})$ (processes 261,262,266,267 \cite{MCFM})
               obtained with PDF including an intrinsic charm component 
	       (CTEQ66c) and PDF having only an extrinsic component (CTEQ66). Heavy flavor jet
	       tagging efficiencies have been applied to the $c$-jets and the $b$-jets. Right: Ratio of these 
	       two spectra.}
\label{Fig_9Zall}
\end{figure}

   In order to be able to observe and quantify the intrinsic charm contribution to the proton, the 
   size of the effect presented in Fig.~\ref{Fig_9Zall} must be larger than the total systematic 
   uncertainty in each bin of the measured heavy flavor jet spectrum. The experimental
   uncertainties on background estimates, jet energy scale and resolution effects, and heavy-quark 
   tagging efficiency are typically large, as can be inferred from the latest ATLAS~\cite{atlZb,atlWb}  
   and CMS~\cite{cmsZb,cmsWb} publications on $Z+b$ and $W+b$ measurements. Recently,
   ATLAS published a measurement of $W$+jets to $Z$+jets differential cross section ratio as a function
   of a plethora of observables~\cite{atlRjets}. The results indicate a substantial reduction of the
   main systematic uncertainties with respect to the absolute differential cross section measurements
   also performed by the ATLAS Collaboration~\cite{atlWjets, atlZjets}. This strategy to offset systematic
   uncertainties could be exploited to measure intrinsic charm contribution to the proton, if the
   sensitivity to intrinsic charm is not washed out by the ratio. 
   
   To test this, similarly as what was done for $Z+Q$, we used MCFM to calculate, at NLO in QCD, the $p_T$ 
   spectra of the leading heavy flavor jets ($b+c$) produced in association with $W^\pm$ boson in 
   hard $pp$ $\sqrt{s}=$8 TeV collisions. The $Wb$, $Wc$, and $Wbj$ contributions of MCFM 
   (processes 12, 13, 14, 17, 18, 401, 402, 406 and 407) have been summed and the $b$-jet and 
   $c$-jet tagging efficiencies have been applied. In all cases, the W-boson is decaying leptonically
   and is required to satisfy $1.52<\eta_W<2.4$. As can be seen in Fig.~\ref{Fig_10Wall}, comparing
   the heavy flavor jet $p_T$ spectra when intrinsic charm is included or not in the PDF (CTEQ66c vs
   CTEQ66), the sensitivity of $W+Q$ to intrinsic charm is small. Taking the ratio of $Z+Q$ to $W+Q$ 
   should therefore not smear out the effects observed in $Z+Q$ alone. To verify this, the ratio of the 
   $p_T$ spectra of the leading heavy flavor jet ($b/c$) produced in $Zb+Zc+Zb(+b)+Zc(+c)$ and 
   $Wb+Wc+Wbj$ processes has been calculated using the PDF with the {\it IC} and without the {\it IC} 
   contribution and is presented in Fig.~\ref{Fig_11Rall}. As can be seen in this figure, the {\it IC} signal 
   observed in $Z+Q$ is not affected by the ratio, which can amount to about 200 $\%$ of the extrinsic 
   only contribution at $p_T^{jet}$ of about 270-300 GeV$/$c. This ratio measurement would likely to, at 
   least, partially cancel a number of large experimental systematic uncertainties, especially since in 
   our proposal, $V+c$-jets and $V+b$-jets are both considered as signal and not treated as a 
   background with respect to the other. This would allow for  a clear signal at the LHC, if the {\it IC}
   contribution is sufficiently high (here we considered a 3.5\% contribution).

\begin{figure}[h!]
\begin{center}
\begin{tabular}{cc}
\mbox{\epsfig{file=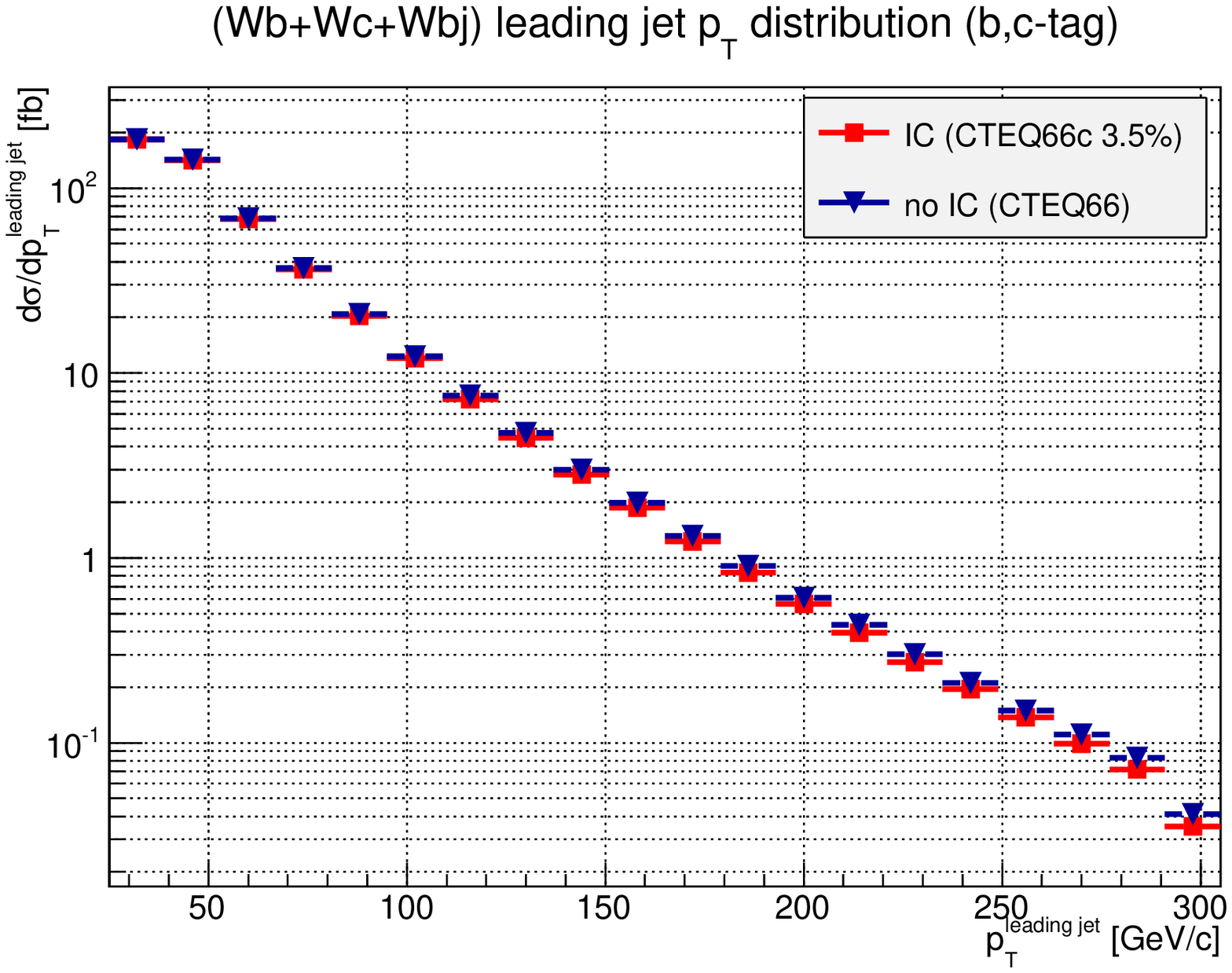,width=0.50\linewidth}
\epsfig{file=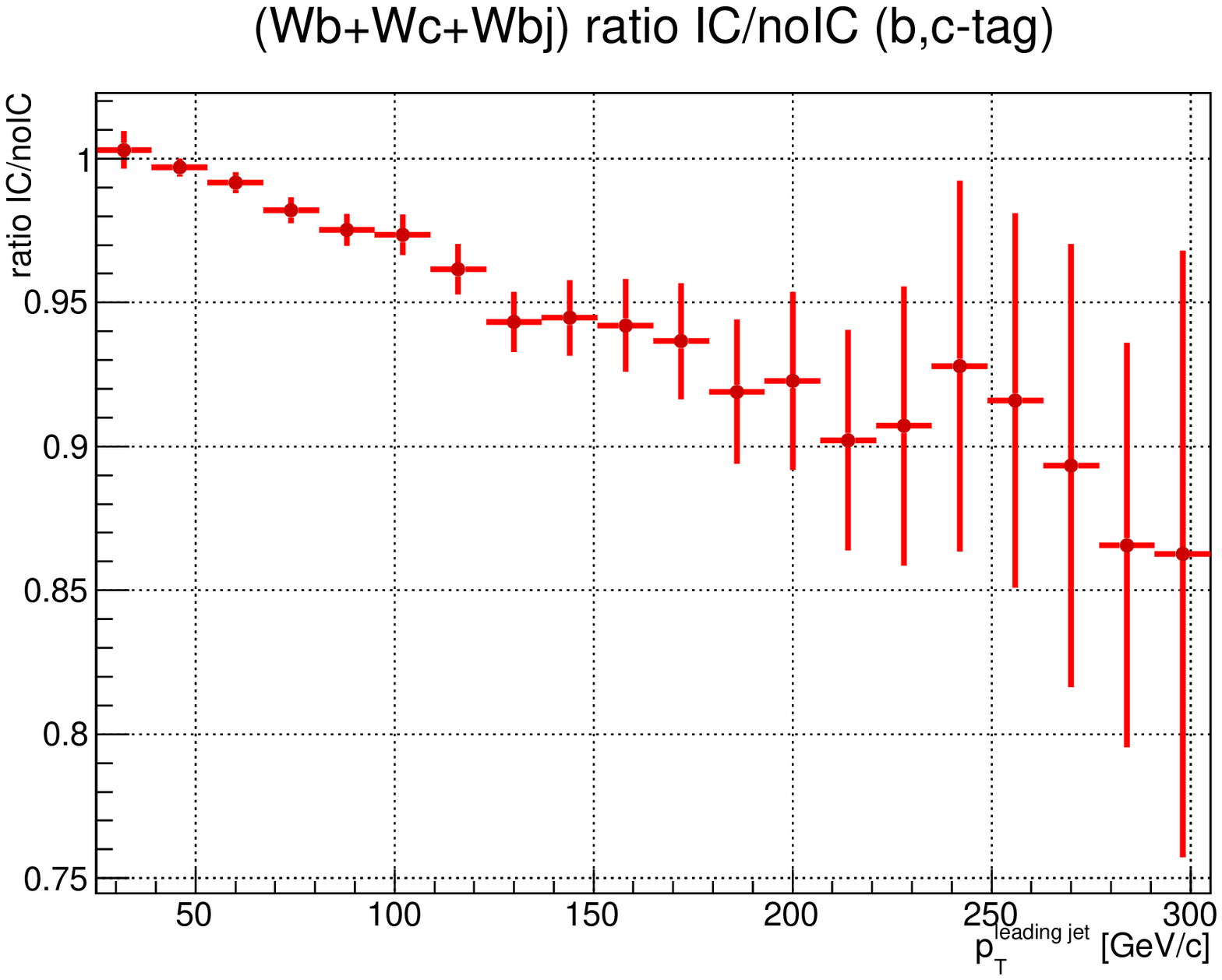,width=0.50\linewidth}}
\end{tabular}
\end{center}
\caption{Left: Comparison of the $p_T$-spectra for the total NLO $pp\rightarrow W+b$ plus  
	       $pp\rightarrow W+c$ plus $pp\rightarrow W+bj$ processes 
               (12,17,13,18,401,402,406,407 \cite{MCFM}) 
               obtained with PDF including an intrinsic   
	       charm component (CTEQ66c) and PDF having only an extrinsic component (CTEQ66). 
	       Heavy flavor jet tagging efficiencies have been applied to the $c$-jets and the $b$-jets. Right: 
	       Ratio of these two spectra.}
\label{Fig_10Wall}
\end{figure} 
\vspace{0.2 cm}

\begin{figure}[h!]
\begin{center}
\begin{tabular}{cc}
\hspace{0.5cm}
\mbox{\epsfig{file=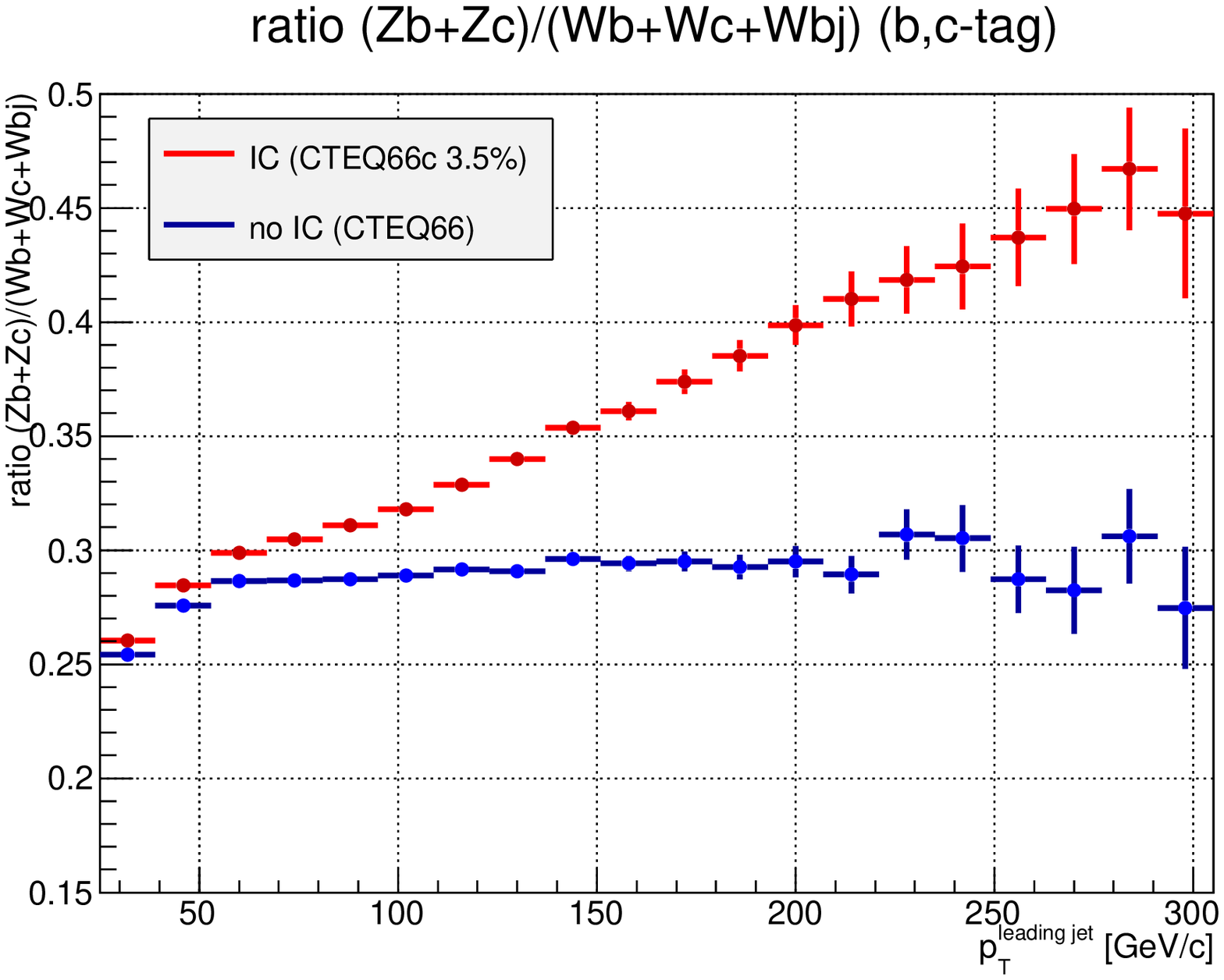,width=0.50\linewidth}
\epsfig{file=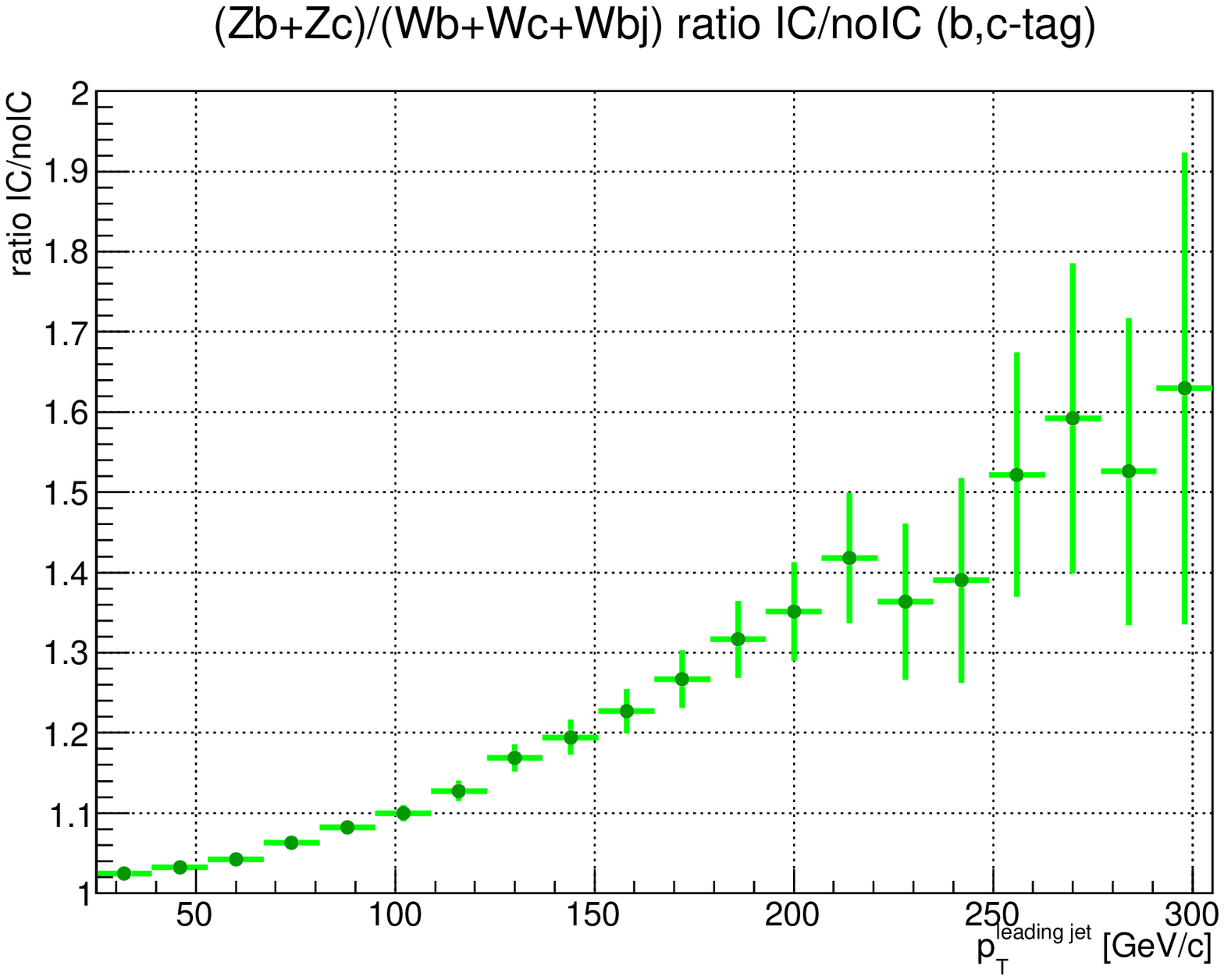,width=0.50\linewidth}}
\end{tabular}
\end{center}
\caption{Left: Comparison of the ratio of the $p_T$-spectra for $Z+Q$ to $W+Q$ NLO  processes obtained with 
	      PDF including an intrinsic charm component (CTEQ66c) and PDF having only an extrinsic 
	      component (CTEQ66). Heavy flavor jet tagging efficiencies have been applied to the $c$-jets 
	      and the $b$-jets. Right:  Ratio of these two ratios of spectra.}
\label{Fig_11Rall}
\end{figure} 

\section{Conclusion}
      In this paper we have shown that the possible existence of an intrinsic heavy quark component 
      to the proton can be seen not only in the forward open heavy flavor production of $pp$-collisions
      (as it was believed before) but it can also be observed in the semi-inclusive $pp$-production of 
      massive vector bosons in association with heavy flavor jets ($b$, and $c$). In particular, 
      it was shown that the {\it IC} contribution can produce much more $Z+c$-jet events (factor 1.5 -- 2) than
      what is predicted from extrinsic contribution to PDF alone, when the heavy flavor jet has a transverse
      momentum of $p^{jet}_{T}>$ 100 GeV$/$c and a pseudo-rapidity satisfying 1.5$<\mid \eta_{jet}\mid<$ 2.4.
      We then showed that this conclusion stays true when the $Z+b$ negative contribution and the inefficiencies in 
      the experimental identification of heavy flavor jets are taken into account. 
     
      We then showed that because of the dominant contribution of gluon-splitting processes, the production 
      of W-bosons accompanied by heavy flavor jets is not sensitive to intrinsic quarks. We took advantage
      of this to propose a promising measurement that reduces the expected systematic uncertainties on the 
      measurement results compared to a differential cross section measurement of the heavy flavor jet spectrum in 
      $Z+Q$ events. The idea is to use the ratio of the leading heavy flavor spectra in inclusive heavy flavor
      $Z+Q$ to $W+Q$ events to verify the predictions about an {\it IC} contribution to the proton. 
      Such measurements can already be made with ATLAS and CMS available data. 
 
\vspace{0.5 cm}
{\bf Acknowledgments }\\
We thank S.J. Brodsky and A.A. Glasov for extremely helpful discussions and recommendations by the study 
of this topic. The authors are grateful to H. Jung, A.V. Lipatov, V.A.M. Radescu, A. Cooper-Sarkar and 
N.P. Zotov for very useful discussions and comments. This research was supported by the Russia RFBR grant 
No. 13-02001060 (Lykasov) and by the USA DOE grant Proposal ID: 0000209063
(Beauchemin).

\begin{footnotesize}

\end{footnotesize}


\begin{thebibliography}{99}
\bibitem{LATTICE}
J.W.Negele {\it et al.,} Nucl.Phys. B [Proc.Suppl.] {\bf 128}, (2004) 170;
W.Schroers, Nucl. Phys. A {\bf 755} (2005) 333.
\bibitem{QCD_anal1}
J.Pumplin, D.R.Stump, J.Huston, H.L.Lai, P.Nadolsky and W.K.Tung,
J. High Energy Physics {\bf 07} (2002) 012;
D.R.Stump, J.Huston. J.Pumplin, W.K.Tung,  H.L.Lai, S.Kuhlmann and J.F.Owens,
J. High Energy Physics {\bf 10} (2003) 046.
\bibitem{QCD_anal2}
R.S.T${\:o}$rne, A.d.Martin, W.G.Stirling, and R,G.Roberts, arXiv:04073 [hep-ph0]. 
\bibitem{DGLAP}
V.N.~Gribov and L.N.~Lipatov, Sov.J.Nucl.Phys. {\bf 15} (1972) 438;
G.~Altarelli and G.~Parisi, Nucl.Phys. B {\bf 126} (1997) 298;
Yu.L.~Dokshitzer, Sov.Phys. JETP {\bf 46} (1977) 641.
\bibitem{atlas-s}
The ATLAS Collaboration, Phys. Rev. Lett. {\bf 109}, 012001 (2012) 

\bibitem{H1:2005}
A.Aktas, {\it etal.,} (H1 Collaboration), Eur.Phys. J.C40,
(2005) 349; arXiv:0507081 [hep-ex].
 
\bibitem{Brodsky:1980pb}
Brodsky S., Hoyer P., Peterson C., Sakai N., 
Phys.Lett. B {\bf 93} (1980) 451. 

\bibitem{Brodsky:1981}
Brodsky S., Peterson C., Sakai N.,
 Phys.Rev. D, {\bf 23} (1981) 2745.


\bibitem{Golowich:1981}
J.F. Donoghue, E. Golowich, Phys.Rev.D {\bf 15} (1977) 3421.

\bibitem{Pumplin:2005yf}
Pumplin J., 
Phys.Rev. D {\bf 73} (2006) 114015.

\bibitem{Peng_Chang:2012}
Jen-Chien Peng, Wen-Chen Chang,
Sixth Intern. Conf., on Quarks and Nuclear Physics,
Apriel 16-20, 2012, Ecole Polytechnique Palaiseau, Paris,
arXiv:1207.2193 [hep-ph].

\bibitem{LBDS:2013}
G.I.Lykasov, I.V.Bednyakov, M.A.Demichev, Yu.Yu.Stepanenko,
Nucl.Phys. B [Proc.Suppl.] 245 (2013) 215.

\bibitem{LBPZ:2012}
G.I.Lykasov, V.A.Bednyakov, A.F.Pikelner and N.I.Zimin,
Eur.Phys.Lett. {\bf 99} (2012) 21002; arXiv:1205.1131v2 [hep-ph].

\bibitem{BDLST:2014}
V.A.Bednyakov, M.A.Demichev, G.I.Lykasov, T.Stavreva, M.Stockton,
Phys.Lett.B {\bf 728} (2014) 602. 

\bibitem{Pumplin:2007wg}
Pumplin J., Lai H., Tung W., 
Phys.Rev. D {\bf 75} (2007) 054029.

\bibitem{Nadolsky:2008zw}
Nadolsky P.~M., {\it et. al.,} 
Phys. Rev. D {\bf 78} (2008) 013004.

\bibitem{Polyakov:1998rb}
Polyakov M.~V., Schafer A., Teryaev O.~V., 
Phys.Rev. D {\bf 60} (1999) 051502.



\bibitem{D0:2009}
V.M.Abazov {\it et al.,} 
Phys.Rev.Lett. {\bf 102} (2009) 192002;
arXiv:0901.0739 [hep-ex].

\bibitem{D0:2012}
V.M.Abazov {\it et al.,} 
Phys.Lett. B {\bf 714} (2012) 32;
arXiv:1203.5865 [hep-ex].

\bibitem{D0:2013}
V.M.Abazov {\it et al.,} 
Phys.Lett. B {\bf 719} (2013) 32;
arXiv:1210.5033 [hep-ex].

\bibitem{CDF:2010}
T.Altonen {\it et al.,} 
Phys.Rev. D81 (2010) 052006; arXiv:0912.3453 [hep-ex].    

%
%
%

\bibitem{Stavreva:2009vi}
   T.P.  Stavreva, and J.F. Owens,
   Phys.Rev. D {\bf 79} (2009) 054017;
   arXiv:0901.3791 [hep-ph].

\bibitem{MCFM}
J.M. Campbell and R.K. Ellis, Phys. Rev. {\bf D} 65 (2002) 113007;
http://mcfm.fnal.gov/ 

\bibitem{ATLAS:13-109}
The ATLAS Collaboration,
ATLAS Note, ATLAS-CONF-2013-109; 21/02/2014;
ATLAS-CONF-2014-046, 06 July 2014.


\bibitem{atlZb}
The ATLAS Collaboration, CERN-PH-EP-2014-118, submitted to JHEP (2014).

\bibitem{atlWb}
The ATLAS Collaboration, JHEP {\bf 06}, 084 (2013).


\bibitem{cmsZb}
The CMS Collaboration, JHEP {\bf 1406}, 120 (2014)  

\bibitem{cmsWb}
The CMS Collaboration, PLB {\bf 735}, 204 (2014) 

\bibitem{atlRjets}
The ATLAS Collaboration, Phys. Lett. B{\bf 708}, 221-240 (2012);
The ATLAS Collaboration, ATLAS-COM-CONF-2014-049, Submitted to EPJC  (2014).


\bibitem{atlWjets}
The ATLAS Collaboration, ATLAS-COM-CONF-2014-050, Submitted to EPJC (2014).

\bibitem{atlZjets}
The ATLAS Collaboration, JHEP {\bf07}, 032 (2013).



%



\end{thebibliography}
\end{document}